\documentclass[10pt]{article}

\providecommand{\tightlist}{%
  \setlength{\itemsep}{0pt}\setlength{\parskip}{0pt}}

\usepackage[margin=0.75in]{geometry}
\usepackage{multicol}
\usepackage{dblfloatfix}  
\usepackage{amsmath,amssymb}
\usepackage{graphicx}
\usepackage[hyphens,spaces,obeyspaces]{url}
\usepackage{hyperref}
\usepackage{xcolor}
\usepackage{fancyhdr}
\usepackage{titlesec}
\usepackage{booktabs}
\usepackage{longtable}
\usepackage{array}
\usepackage{calc}
\usepackage{caption}
\usepackage{enumitem}
\usepackage{fvextra}  
\usepackage{framed}
\usepackage{listings}
\usepackage[export]{adjustbox}  

\usepackage{float}
\makeatletter
\newenvironment{inlinefigure}[1][htbp]{%
  \par\addvspace{\abovecaptionskip}%
  \noindent\begin{minipage}{\columnwidth}%
  \def\@captype{figure}%
  \centering%
}{%
  \end{minipage}%
  \par\addvspace{1.5em}%
}
\makeatother

\renewenvironment{figure}[1][htbp]{%
  \inlinefigure[#1]%
}{%
  \endinlinefigure%
}

\newcommand{\pandocbounded}[1]{%
  \adjustbox{max width=\columnwidth}{#1}%
}

\newcounter{none}
\definecolor{shadecolor}{RGB}{248,248,248}

\let\oldtexttt\texttt
\renewcommand{\texttt}[1]{%
  {\small\oldtexttt{#1}}%
}

\newenvironment{Shaded}{%
  \begin{snugshade}%
  \footnotesize%
  \setlength{\parindent}{0pt}%
  \setlength{\linewidth}{\columnwidth}%
  \setlength{\hsize}{\columnwidth}%
  \raggedright%
  \par%
}{%
  \par%
  \end{snugshade}%
}



\setlist{nosep, leftmargin=1.5em, topsep=0.5em}
\setlist[itemize]{label=\textbullet}
\setlist[enumerate]{label=\arabic*.}

\newcommand{\CommentTok}[1]{\textcolor[rgb]{0.38,0.63,0.69}{\textit{#1}}}

\newcommand{\ControlFlowTok}[1]{\textcolor[rgb]{0.00,0.44,0.13}{\textbf{#1}}}
\newcommand{\DataTypeTok}[1]{\textcolor[rgb]{0.56,0.13,0.00}{#1}}
\newcommand{\DecValTok}[1]{\textcolor[rgb]{0.25,0.63,0.44}{#1}}

\newcommand{\NormalTok}[1]{#1}
\newcommand{\OperatorTok}[1]{\textcolor[rgb]{0.40,0.40,0.40}{#1}}

\DefineVerbatimEnvironment{Highlighting}{Verbatim}{%
  commandchars=\\\{\},
  fontsize=\footnotesize,
  breaklines=true,
  breakanywhere=true,
  breaksymbol=,
  breakautoindent=false
}

\let\oldShaded\Shaded
\let\oldendShaded\endShaded
\renewenvironment{Shaded}{%
  \begin{minipage}{\columnwidth}%
  \oldShaded%
}{%
  \oldendShaded%
  \end{minipage}%
}

\hypersetup{
    colorlinks=true,
    linkcolor=blue,
    citecolor=blue,
    urlcolor=blue,
    breaklinks=true,
}

\titleformat{\section}
  {\normalfont\large\bfseries}{\thesection}{1em}{}
\titleformat{\subsection}
  {\normalfont\normalsize\bfseries}{\thesubsection}{1em}{}
\titleformat{\subsubsection}
  {\normalfont\small\bfseries}{\thesubsubsection}{1em}{}

\makeatletter
\renewcommand{\maketitle}{%
\begin{center}%
{\LARGE\bfseries LLM Agent-Assisted Reverse Engineering with
Quantitative Readability Metrics\par}%
\vspace{0.5em}%
{\large Neil Archibald\par}%
{\large Ruben Thijssen\par}%
{\normalsize\itshape Commonwealth Bank of Australia\par}%
{\small\texttt{neil.archibald@cba.com.au,
ruben.thijssen@cba.com.au}\par}%
\vspace{0.5em}%
{\normalsize May 2026\par}%
\end{center}%
\begin{quote}%
\small\textbf{Abstract:} Automatic decompilers produce functionally
correct but often unreadable C code. This paper addresses one stage of
the reverse engineering workflow: improving the readability of
decompiled code using LLM agents guided by quantitative metrics. We
present a three-phase research evolution. Phase 1 (tool-driven steering
via Ghidra MCP) suffered from incomplete coverage and inconsistent
improvements due to lack of quantitative guidance. Phase 2 (structural
similarity validation alone) revealed that agents optimize for metrics
in unintended ways, producing structurally equivalent but less readable
code. Our contribution is the \textbf{Quantitative Readability Score
(QRS)} framework, a composite metric combining a structural similarity
gate with three independent readability sub-metrics (Lexical Surprisal,
Structural Simplicity, and Idiomatic Quality). We demonstrate that
QRS-guided refinement enables LLM agents to make targeted readability
improvements without sacrificing correctness. We provide a discussion of
the broader reverse engineering workflow (binary lifting, decompilation
cleanup, and achieving functional equivalence) as context, however, it
remains out of scope.%
\end{quote}%
\noindent\small\textbf{Keywords:} binary decompilation, code readability
metrics, LLM agents, reverse engineering, quantitative rewards\par%
\vspace{1em}%
}
\makeatother

\begin{document}

\maketitle

\begin{multicols}{2}

\section{Introduction}\label{introduction}

\subsection{Motivation}\label{motivation}

Binary reverse engineering is essential for security analysis,
vulnerability research, legacy software maintenance, and malware
analysis. Automatic decompilation tools produce code that is
functionally correct but difficult to read. Variable names like
\texttt{local\_48} - which communicate storage context rather than
semantic meaning, unnecessarily complex control flow, and pervasive
low-level patterns - created by compiler optimizatons, obscure the
developer's intent.

Large Language Models (LLMs) can write, read, and reason about code,
making them natural candidates for improving decompiler output. However,
without quantitative guidance on what constitutes an improvement, agents
exhibit recurring failure modes: they lose focus across iterations
(context rot), optimize single metrics in unintended ways
(gamification), introduce subtle functional regressions, and produce
inconsistent results due to their non-deterministic nature. These
problems compound in long-running tasks where the agent must maintain
coherence across many functions and iterations.

\subsection{Problem Statement}\label{problem-statement}

Source code produced by automated decompilers is widely regarded as hard
to read regardless of the specific tool used. This creates demand for
highly skilled analysts in teams performing malware analysis, exploit
development, and patch analysis. To use LLMs effectively for this work,
we need to:

\begin{enumerate}
\def\labelenumi{\arabic{enumi}.}
\tightlist
\item
  \textbf{Measure multiple dimensions} of readability to avoid
  single-metric gamification
\item
  \textbf{Provide interpretable feedback} so agents know specifically
  what to improve
\item
  \textbf{Validate correctness} by measuring structural similarity of
  the recompiled binary
\item
  \textbf{Scale to long-running refinement} across multiple functions
  without manual intervention
\end{enumerate}

\subsection{Research Evolution}\label{research-evolution}

This paper describes a three-phase progression toward metric-guided
agent refinement:

\begin{itemize}
\item
  \textbf{Phase 1 (Ghidra MCP)}: Direct agent control of Ghidra
  decompilation. Incomplete coverage, inconsistent quality, and no
  feedback loop.
\item
  \textbf{Phase 2 (radare2 CFG comparison)}: Agent loop maximizing
  structural similarity. Functional equivalence achieved, but
  readability degraded as agents gamified the single metric.
\item
  \textbf{Phase 3 (QRS-Guided Refinement)}: Readability refinement
  guided by a composite metric with continuous binary validation.
  Targeted, balanced improvements while maintaining correctness.
\end{itemize}

Each phase informed the next. The broader decompilation workflow (binary
lifting, control flow recovery, type inference, and producing an initial
compilable decompilation) is a prerequisite to QRS-guided refinement but
not the focus of this paper. QRS assumes a compilable decompilation is
already available.

\subsection{Paper Structure}\label{paper-structure}

\begin{itemize}
\tightlist
\item
  \textbf{Section 2} describes our tool-driven approach and its
  limitations
\item
  \textbf{Section 3} explains the structural similarity validation
  strategy and the metric gaming problem
\item
  \textbf{Section 4} presents the QRS framework and its three
  readability sub-metrics
\item
  \textbf{Section 5} presents experimental results across both QRS
  experiments
\item
  \textbf{Section 6} discusses future work, including the out-of-scope
  decompilation cleanup problem and extension opportunities to the QRS
  framework
\item
  \textbf{Section 7} concludes with implications for agentic reverse
  engineering
\end{itemize}

\section{Phase 1: Tool-Driven Steering with Ghidra
MCP}\label{phase-1-tool-driven-steering-with-ghidra-mcp}

\subsection{Approach}\label{approach}

Our initial approach gave an LLM agent direct control of Ghidra
(\url{https://www.nsa.gov/ghidra}) via a Model Context Protocol (MCP)
server. By exposing the extensive list of Ghidra APIs as MCP tools, the
agent could load a binary, iteratively disassemble and decompile
functions, rename symbols based on its analysis, refine types and data
structures, and gradually produce more readable output.

\emph{Note: Rather than building a custom MCP server, we used the
existing open-source implementation at:
\url{https://github.com/LaurieWired/GhidraMCP}}

This design gave the agent full autonomy over the analysis process. It
decided which functions to prioritize, where renaming would add clarity,
and how to sequence its work. We steered the agent toward
disassembly-level analysis rather than relying heavily on Ghidra's
decompiler output, to avoid decompilation errors tainting downstream
decisions. The agent pushed updates back to the Ghidra database,
producing a usable project file for post-agent analysis.

\subsection{Observed Failures}\label{observed-failures}

In practice, Phase 1 failed in several ways.

The agent exhibited \textbf{incomplete coverage}. It would decompile the
first few functions with high effort, but quickly would start producing
stub implementations. In binaries with hundreds of functions, only
10--15\% received genuine analysis.

Improvements were \textbf{inconsistent}. Without a quantitative
definition of ``better,'' the agent made ad-hoc decisions: renaming a
function based on a single cross-reference, or leaving it untouched when
uncertain. Type assignments were sporadic and often incorrect.

The agent also suffered from \textbf{context loss} (or also known as
\textbf{context rot}) across functions. After improving Function A and
moving to Function B, it would later misremember Function A's signature
when examining a caller, producing inconsistent naming and typing across
the binary.

Finally, there was \textbf{no feedback loop}. When the agent produced an
output, nothing told it whether the result was an improvement. Without
measurement, it could not learn which decisions generalized and which
were noise.

\subsection{Analysis}\label{analysis}

The root cause was the absence of quantitative guidance. The agent had
tools (Ghidra functions) and an implicit objective (improve
readability), but no way to measure progress. Without a reward signal,
it reverted to local heuristics, reduced scope through shortcuts, and
eventually refused to continue.

This demonstrated that tool access alone is insufficient for
long-horizon reverse engineering tasks. Agents need explicit, measurable
objectives to maintain consistency across many functions and iterations.
Phase 2 introduced such a metric: structural similarity.

\section{Phase 2: Binary Equivalence
Validation}\label{phase-2-binary-equivalence-validation}

\subsection{Approach}\label{approach-1}

Phase 1 failed because the agent had no quantitative feedback. To
address this, we introduced structural similarity as a reward signal: a
decompiled function is considered correct if it recompiles to machine
code that is structurally similar to the original.

We evaluated true semantic equivalence approaches, including Ghidra's
P-code lifting and angr-based symbolic execution. These move closer to
behavioural validation but introduced significant computational cost and
engineering complexity without sufficient practical improvement. We
opted instead for a structurally grounded, computationally tractable
proxy.

We implemented an MCP server that accepts two binaries (original and
agent-generated), performs function-level CFG comparison using radare2,
and returns similarity scores (0.0--1.0) per function and in aggregate.
We chose radare2 over more advanced tools like BinDiff or GhidraDiff for
its low overhead, which allowed rapid iteration during agent refinement.

\subsubsection{Similarity Methodology}\label{similarity-methodology}

Our implementation measures \textbf{structural similarity}, not true
\textbf{semantic equivalence}. The primary mechanism uses radare2's
native control-flow graph comparison (\texttt{cg}) to evaluate CFG
similarity between corresponding functions.

We employ a hybrid scoring strategy:

\begin{itemize}
\tightlist
\item
  \textbf{Primary:} Native CFG comparison via radare2's \texttt{cg}
  command
\item
  \textbf{Fallback} (when native diffing fails): Multi-dimensional
  scoring combining size similarity (50\%), basic block similarity
  (30\%), and instruction similarity (20\%)
\end{itemize}

All metrics are structural. They compare binary layout and control-flow
characteristics rather than runtime behaviour.

True semantic equivalence generally requires symbolic execution or
exhaustive differential testing across all possible inputs. These
approaches are computationally intractable in the general case. Two
programs may be semantically equivalent while exhibiting low structural
similarity due to different compilation strategies. Conversely, high
structural similarity does not guarantee identical behaviour in edge
cases. Structural similarity is a pragmatic compromise: computationally
feasible and effective at detecting most unintended behavioural
deviations, even if it cannot formally guarantee equivalence.

\subsubsection{Iteration Architecture}\label{iteration-architecture}

Our research found that shorter, context-constrained LLM executions
produce higher quality output than large monolithic sessions with
continuously expanding context windows. Agents perform better when
prompted with a clear, achievable objective (a micro-prompt) and told
they are part of an iterative process. This prevents moonshot
approaches. By requesting a ``small iterative improvement,'' the agent
regularly considers its task complete, creating natural restart points
where a new agent receives an updated prompt with specific guidance.
This produces a threshold-driven workflow from non-deterministic LLM
tasks, eliminating the need for monolithic error-recovery processes.

\subsubsection{Implementation}\label{implementation}

The radare2 CFG comparison MCP (r2dmcp) exposed the following tools:

\begin{itemize}
\tightlist
\item
  \texttt{compare\_binaries}: Compare two binary files and return
  function similarity scores
\item
  \texttt{list\_functions}: List all functions in a binary file
\item
  \texttt{compare\_functions}: Compare specific functions between two
  binaries
\item
  \texttt{get\_diff\_summary}: Get a summary of differences between two
  binaries
\end{itemize}

The agent loop operated as follows:

\textbf{Algorithm 1: Binary Similarity Optimization}

\begin{verbatim}

Input: Original binary B
Output: Refined C code with high similarity

1. C = decompile(B)
2. s = similarity(compile(C), B)
   // typically 0.5-0.7
3. while s < threshold do
4.     scores = function_similarities(C, B)
5.     targets = low_similarity_funcs(scores)
6.     C' = agent_refine(C, scores, targets)
7.     s' = similarity(compile(C'), B)
8.     if s' > s then
9.         C = C'
10.        s = s'
11. end if
12. end while
13. return C
\end{verbatim}

\subsection{The Gamification Problem}\label{the-gamification-problem}

The similarity loop initially worked. Scores climbed from 0.5--0.7 to
0.85--0.95 within a few iterations. However, examining the ``improved''
code revealed a problem: readability was degrading.

Code that had been partially reverse-engineered into readable form was
regressing toward low-level complexity:

\begin{Shaded}
\begin{Highlighting}[]
\CommentTok{// Before refinement (messy but readable)}
\DataTypeTok{int64\_t}\NormalTok{ compute\_sum}\OperatorTok{(}\DataTypeTok{int}\OperatorTok{*}\NormalTok{ arr}\OperatorTok{,} \DataTypeTok{int}\NormalTok{ len}\OperatorTok{)} \OperatorTok{\{}
    \DataTypeTok{int64\_t}\NormalTok{ result }\OperatorTok{=} \DecValTok{0}\OperatorTok{;}
    \ControlFlowTok{for} \OperatorTok{(}\DataTypeTok{int}\NormalTok{ i }\OperatorTok{=} \DecValTok{0}\OperatorTok{;}\NormalTok{ i }\OperatorTok{\textless{}}\NormalTok{ len}\OperatorTok{;}\NormalTok{ i}\OperatorTok{++)} \OperatorTok{\{}
\NormalTok{        result }\OperatorTok{+=}\NormalTok{ arr}\OperatorTok{[}\NormalTok{i}\OperatorTok{];}
    \OperatorTok{\}}
    \ControlFlowTok{return}\NormalTok{ result}\OperatorTok{;}
\OperatorTok{\}}

\CommentTok{// After agent refinement (high similarity, unreadable)}
\DataTypeTok{int64\_t}\NormalTok{ compute\_sum}\OperatorTok{(}\DataTypeTok{int}\OperatorTok{*}\NormalTok{ arr}\OperatorTok{,} \DataTypeTok{int}\NormalTok{ len}\OperatorTok{)} \OperatorTok{\{}
    \DataTypeTok{int64\_t}\NormalTok{ rax }\OperatorTok{=} \DecValTok{0}\OperatorTok{;}
    \ControlFlowTok{for} \OperatorTok{(}\DataTypeTok{int64\_t}\NormalTok{ rcx }\OperatorTok{=} \DecValTok{0}\OperatorTok{;}\NormalTok{ rcx }\OperatorTok{\textless{}}\NormalTok{ len}\OperatorTok{;}\NormalTok{ rcx}\OperatorTok{++)} \OperatorTok{\{}
\NormalTok{        rax }\OperatorTok{+=} \OperatorTok{*(}\NormalTok{arr }\OperatorTok{+}\NormalTok{ rcx }\OperatorTok{*} \DecValTok{8}\OperatorTok{);}
    \OperatorTok{\}}
    \ControlFlowTok{return}\NormalTok{ rax}\OperatorTok{;}
\OperatorTok{\}}
\end{Highlighting}
\end{Shaded}

The agent had discovered that the easiest path to high similarity was to
minimize changes: keep decompiler variable names (\texttt{var\_48},
\texttt{local\_8}), avoid type inference (use void pointers), manually
implement stack cookies, and preserve unusual control flow. The less it
changed, the more similar the output. This is rational behaviour for an
agent optimizing a single metric. When a measure becomes a target, it
ceases to be a good measure.

\subsection{Analysis}\label{analysis-1}

The failure stemmed from a mismatch between what we measured and what we
wanted. Binary similarity measures structural fidelity. But reverse
engineering aims to produce code a human can understand. A function can
be structurally correct (high similarity) and still unreadable due to
poor naming, decompiler artifacts, and unnecessary low-level patterns.

By optimizing exclusively for structural similarity, we defined success
as mechanical fidelity. The agent maximized the only signal available.
Because readability was not part of the reward function, it was not
improved. In many cases it was actively harmed.

The lesson is that single-metric optimization is insufficient for
multi-objective tasks. Reverse engineering requires functional
correctness, readability, clear naming, proper abstraction, and internal
consistency. No single structural metric encodes all of these. Phase 3
addressed this by designing a multidimensional evaluation system that
preserves correctness while explicitly measuring readability.

\section{The Quantitative Readability Score (QRS)
Framework}\label{the-quantitative-readability-score-qrs-framework}

The Quantitative Readability Score (QRS) is a metric framework designed
to evaluate and guide improvements in source code readability,
specifically focused on improving source code generated by a decompiler
like Ghidra, IDA, or RetDec.

QRS was designed using the following principles:

\begin{enumerate}
\def\labelenumi{\arabic{enumi}.}
\tightlist
\item
  \textbf{Independence}: Sub-metrics must be calculated independently
  from each other, and must be normalised (0.0-1.0 scale) so they can be
  weighted. \emph{Using a weighted formula also allows QRS to be easily
  extended.}
\item
  \textbf{Actionable}: Sub-metrics must have a clear scope and provide
  actionable feedback to an agent on how to improve.
\item
  \textbf{Gaming-resistant}: Metrics should be designed to identify
  specific readability violations (e.g.~decompiler artifacts, complex
  control flow, non-idiomatic patterns) and avoid rewarding code that
  contains preferred patterns. This is critical to prevent gaming and
  ensure that improvements are meaningful. Additionally, balanced
  sub-metric weightings reduce the risk of gamification as they force
  agents to improve 2 or more sub-metrics to achieve the QRS threshold.
  Lastly, we must assume that the model can (or will) know how each
  sub-metric is calculated as they become more capable (e.g.~they
  consume this paper).
\item
  \textbf{Deterministic and Efficient}: Each metric is deterministic and
  computationally efficient, enabling rapid feedback between iterations.
\item
  \textbf{Robustness}: The overall metric should not only measure code
  readability but also protect the workflow from diverging from the
  original objective across multiple iterations due to LLM
  hallucinations, context-rot, or other common issues experienced in
  long-running LLM processes.
\end{enumerate}

\subsection{Architecture: Structural Similarity Gate + Three Readability
Sub-Metrics}\label{architecture-structural-similarity-gate-three-readability-sub-metrics}

QRS is a single metric incorporating these design principles so we can
measure a \textbf{structural similarity quality gate} and three
\textbf{source code readability sub-metrics}:

\[
\text{QRS} = BS \times C
\]

where \(BS \in \{0.1, 1\}\) is the structural similarity gate (radare2
CFG comparison) and \(C\) is the composite readability score:

\[
\begin{aligned}
C = &\;0.34 \text{LS} + 0.33 \text{SS} + 0.33 \text{IQ}
\end{aligned}
\]

where:

\begin{itemize}
\tightlist
\item
  LS=Lexical Surprisal (Section 4.3)
\item
  SS=Structural Simplicity (Section 4.4)
\item
  IQ=Idiomatic Quality (Section 4.5)
\end{itemize}

\textbf{Weighting Rationale}:

The weightings were determined based on a combination of empirical
testing and theoretical considerations. They are intended to balance the
importance of each metric equally:

\begin{itemize}
\tightlist
\item
  LS (0.34): Code should use (statistically) common and familiar idioms
  and patterns
\item
  SS (0.33): Control flow complexity directly impacts readability and
  should avoid deep nesting and long functions
\item
  IQ (0.33): Source code should use clear variable/function names with
  appropriate data types, and avoid goto statements, hex literals,
  pointer arithmetic, and manual memory management.
\end{itemize}

While these weightings proved effective during our experiments, they
should be considered as guidance, not as fixed/absolute values. We
suspect they will need to be adjusted outside of an academic setting as
source code readability is subjective and linked to the quality of the
original source code.

\subsection{Structural Similarity
Gate}\label{structural-similarity-gate}

\begin{verbatim}
function StructuralSimilarityGate(binary_similarity)
Input: binary_similarity from radare2 CFG comparison
Output: gate value in {0.1, 1.0}

1. if binary_similarity >= 0.85 then
2.     return 1.0  // Allow QRS score
3. else
4.     return 0.1  // Reject: functionally incorrect
5. end if
\end{verbatim}

This binary equivalence gate penalises agents for producing source code
that fails to compile or diverges (15\%+) from the original binary. This
is critical and prevents agents from over-optimising for readability at
the cost of the original functionality.

\textbf{Note:} \emph{The threshold of 0.85 was determined based on
empirical testing and includes a buffer for functional differences
caused by, e.g., compiler optimisations or security features (e.g.~stack
canaries).}

\textbf{Note:} \emph{The default value of 0.1 was used to avoid
multiplying \(C\) by zero. Most decompiler-generated source code does
not compile. Setting \(C\) to 0.0 would not result in an improvement
signal after a single iteration, even if the agent made improvements to
any of the other sub-metrics, but was unable to achieve source-code that
compiles. It would force the agent to only focus on achieving semantic
equivalence without ever being rewarded for improving any of the other
sub-metrics.}

\textbf{Limitations}: The structural similarity gate does not measure
true semantic equivalence. Two programs may be semantically equivalent
while exhibiting different control-flow structure due to compiler
optimisations, instruction scheduling, or register allocation choices.
Conversely, high structural similarity does not formally guarantee
identical behaviour across all inputs. We adopt structural similarity as
a computationally tractable proxy: it is efficient to compute,
deterministic, and effective at detecting most unintended behavioural
deviations introduced during readability refinement. For the class of
transformations agents typically make (renaming, refactoring,
simplifying control flow), structural similarity reliably detects
functional regressions. Cases where structural divergence does not imply
semantic divergence (e.g., loop unrolling, constant propagation) are
acknowledged as a limitation of this approach.

\subsection{Sub-Metric 1: Lexical Surprisal
(LS)}\label{sub-metric-1-lexical-surprisal-ls}

\textbf{Concept}: Code written with familiar coding patterns and idioms
is more readable than code containing unexpected constructs. We measure
this using \textbf{surprisal} (negative log-likelihood or NLL), which
quantifies the ``surprise'' of a token within its broader context. Lower
surprisal indicates more idiomatic/familiar code patterns.

\textbf{Mathematical Definition}:

\[
\text{LS} = 1 - \frac{s_{\text{avg}}}{s_{\text{max}}}
\]

where \(s_{\text{avg}}\) is the average surprisal per token and
\(s_{\text{max}} = 10.0\) is the maximum expected surprisal. Result is
clipped to \([0, 1]\). Higher LS scores indicate more familiar,
idiomatic code.

\textbf{Preprocessing Requirements}: To ensure LS measures code
structure independently of documentation practices, all code undergoes
preprocessing before evaluation:

\begin{enumerate}
\def\labelenumi{\arabic{enumi}.}
\tightlist
\item
  \textbf{Comment stripping}: Remove all \texttt{//} single-line and
  \texttt{/*\ */} multi-line comments
\item
  \textbf{Preprocessor directive removal}: Skip \texttt{\#include},
  \texttt{\#define}, \texttt{\#pragma}, and other preprocessor
  statements
\item
  \textbf{Function-level extraction}: Evaluate individual function
  bodies rather than entire files
\end{enumerate}

This preprocessing is essential because comments introduce natural
language tokens that significantly increase surprisal. Our controlled
experiments show comments significantly inflate surprisal scores.
Without preprocessing, well-documented code would be unfairly penalised.
This approach is analogous to standard practices in other code quality
metrics (e.g., linters normalise whitespace).

\textbf{Implementation Details}:

\begin{itemize}
\tightlist
\item
  Uses \texttt{Qwen2.5-Coder-0.5B}. Other models trained for code
  generation were not tested, but we expect similar (or better) results,
  especially with larger language models specifically trained on code.
\item
  Computes average NLL (cross-entropy loss) over tokens on preprocessed
  code
\item
  Typical range: 0-10 for C code (lower = more familiar)
\item
  Normalised using inverse linear scaling where lower surprisal yields
  higher scores
\end{itemize}

We observed an accuracy improvement when using larger code-trained
models; however, this gain is counterbalanced by the increased
computational cost associated with larger models.

\textbf{Key Thresholds}:

\begin{itemize}
\tightlist
\item
  LS \textgreater{} 0.75: Code uses familiar patterns
\item
  LS 0.50-0.75: Mix of familiar and unusual patterns
\item
  LS \textless{} 0.50: Significant use of decompiler artifacts or
  anti-patterns
\end{itemize}

\textbf{Gaming Resistance}:

Difficult to game, as it requires writing code that a language model
finds familiar, which naturally correlates with human-readable code.

\textbf{Example} (shown after preprocessing, comments stripped):

High LS (familiar patterns):

\begin{Shaded}
\begin{Highlighting}[]
\DataTypeTok{int}\NormalTok{ find\_max}\OperatorTok{(}\DataTypeTok{int}\OperatorTok{*}\NormalTok{ arr}\OperatorTok{,} \DataTypeTok{int}\NormalTok{ n}\OperatorTok{)} \OperatorTok{\{}
    \DataTypeTok{int}\NormalTok{ max }\OperatorTok{=}\NormalTok{ arr}\OperatorTok{[}\DecValTok{0}\OperatorTok{];}
    \ControlFlowTok{for} \OperatorTok{(}\DataTypeTok{int}\NormalTok{ i }\OperatorTok{=} \DecValTok{1}\OperatorTok{;}\NormalTok{ i }\OperatorTok{\textless{}}\NormalTok{ n}\OperatorTok{;}\NormalTok{ i}\OperatorTok{++)} \OperatorTok{\{}
        \ControlFlowTok{if} \OperatorTok{(}\NormalTok{arr}\OperatorTok{[}\NormalTok{i}\OperatorTok{]} \OperatorTok{\textgreater{}}\NormalTok{ max}\OperatorTok{)}\NormalTok{ max }\OperatorTok{=}\NormalTok{ arr}\OperatorTok{[}\NormalTok{i}\OperatorTok{];}
    \OperatorTok{\}}
    \ControlFlowTok{return}\NormalTok{ max}\OperatorTok{;}
\OperatorTok{\}}
\end{Highlighting}
\end{Shaded}

Low LS (decompiler artifacts):

\begin{Shaded}
\begin{Highlighting}[]
\DataTypeTok{int32\_t}\NormalTok{ find\_max}\OperatorTok{(}\DataTypeTok{int32\_t} \OperatorTok{*}\NormalTok{param\_1}\OperatorTok{,} \DataTypeTok{int32\_t}\NormalTok{ param\_2}\OperatorTok{)} \OperatorTok{\{}
    \DataTypeTok{int32\_t}\NormalTok{ iVar1}\OperatorTok{;}
    \DataTypeTok{int32\_t}\NormalTok{ iVar2 }\OperatorTok{=} \OperatorTok{*}\NormalTok{param\_1}\OperatorTok{;}
    \ControlFlowTok{for} \OperatorTok{(}\NormalTok{iVar1 }\OperatorTok{=} \DecValTok{1}\OperatorTok{;}\NormalTok{ iVar1 }\OperatorTok{\textless{}}\NormalTok{ param\_2}\OperatorTok{;}\NormalTok{ iVar1 }\OperatorTok{=}\NormalTok{ iVar1 }\OperatorTok{+} \DecValTok{1}\OperatorTok{)} \OperatorTok{\{}
        \ControlFlowTok{if} \OperatorTok{(}\NormalTok{param\_1}\OperatorTok{[}\NormalTok{iVar1}\OperatorTok{]} \OperatorTok{\textgreater{}}\NormalTok{ iVar2}\OperatorTok{)} \OperatorTok{\{}
\NormalTok{            iVar2 }\OperatorTok{=}\NormalTok{ param\_1}\OperatorTok{[}\NormalTok{iVar1}\OperatorTok{];}
        \OperatorTok{\}}
    \OperatorTok{\}}
    \ControlFlowTok{return}\NormalTok{ iVar2}\OperatorTok{;}
\OperatorTok{\}}
\end{Highlighting}
\end{Shaded}

\textbf{Note:} \emph{Both examples have already been preprocessed, so
comments and preprocessor directives are not present.}

The high LS example uses familiar variable names (\texttt{max},
\texttt{arr}, \texttt{n}) and standard idioms
(\texttt{for\ (int\ i\ =\ 1;\ i\ \textless{}\ n;\ i++)}), while the low
LS example contains decompiler artifacts (\texttt{param\_1},
\texttt{iVar1}, \texttt{iVar1\ =\ iVar1\ +\ 1} instead of
\texttt{iVar1++}).

\subsection{Sub-Metric 2: Structural Simplicity
(SS)}\label{sub-metric-2-structural-simplicity-ss}

\textbf{Concept}: Simpler control flow and function structure improve
comprehension. We measure three components:

\begin{enumerate}
\def\labelenumi{\arabic{enumi}.}
\tightlist
\item
  Cyclomatic complexity (cc): the number of independent control-flow
  paths induced by branching constructs (e.g.~if, switch, loops)
\item
  Nesting depth (nd): the depth of nested control structures
  (e.g.~nested ifs, loops)
\item
  Function length (fl): the total number of lines of code (LOC) in the
  function
\end{enumerate}

\textbf{Mathematical Definition}:

\[
\text{SS} = 1 - (W_{cc} \cdot n_{cc} + W_{nd} \cdot n_{nd} + W_{fl} \cdot n_{fl})
\]

where \(W_{cc} = 0.50\), \(W_{nd} = 0.30\), \(W_{fl} = 0.20\), and:

\begin{itemize}
\tightlist
\item
  \(n_{cc} = \min(\text{cc} / 20, 1.0)\)
\item
  \(n_{nd} = \min(\text{nd} / 8, 1.0)\)
\item
  \(n_{fl} = \min(\text{fl} / 200, 1.0)\)
\end{itemize}

\textbf{Thresholds and Rationale}:

\begin{itemize}
\tightlist
\item
  Cyclomatic complexity \textless= 20: Research shows complexity
  \textgreater{} 10-15 significantly reduces comprehension
\item
  Max nesting \textless= 8: Deeply nested code requires mental stack
  frames
\item
  Function length (fl) \textless= 200 LOC: Industry standard; functions
  should fit on screen
\end{itemize}

\textbf{Implementation}:

\begin{itemize}
\tightlist
\item
  Uses the open-source \texttt{lizard} library
  (\url{https://github.com/terryyin/lizard}) to compute cyclomatic
  complexity
\item
  Custom AST parsing for nesting depth and line counting
\item
  Applies incremental penalties (not hard limits)
\end{itemize}

\textbf{Key Thresholds}:

\begin{itemize}
\tightlist
\item
  SS \textgreater{} 0.75: Simple, well-structured code
\item
  SS 0.50-0.75: Moderate complexity
\item
  SS \textless{} 0.50: Complex, hard to understand
\end{itemize}

\textbf{Gaming Resistance}: Difficult to game, as reducing complexity
requires actual refactoring, which correlates with readability.

\subsection{Sub-Metric 3: Idiomatic Quality
(IQ)}\label{sub-metric-3-idiomatic-quality-iq}

\textbf{Concept}: Reducing decompiler code patterns
(e.g.~variable/function names, hexadecimal literals, goto statements)
improves code readability. We measure this by calculating the violation
density:

\textbf{Mathematical Definition}:

\[
\text{IQ} = 1 - \frac{\text{violations}}{\text{lines\_of\_code}}
\]

\textbf{Implementation}:

We used \texttt{clang-tidy} (a widely used C/C++ linter) to detect
idiomatic violations in the decompiled code. \texttt{clang-tidy} comes
with a large built-in library of reliable tests. It also provides a
powerful framework for developing custom checks. The following built-in
clang-tidy checks were used to detect idiomatic violations:

\begin{enumerate}
\def\labelenumi{\arabic{enumi}.}
\tightlist
\item
  \textbf{readability-identifier-naming}: Checks for function/variable
  names (e.g.~v1, a2, sub\_401000)
\item
  \textbf{readability-identifier-length}: Checks for very short names
  (1-2 chars)
\item
  \textbf{readability-isolate-declaration}: Identifies when multiple
  vars appear on one line (e.g.~\texttt{int\ x,\ y,\ z})
\item
  \textbf{readability-misleading-indentation}: Finds
  inconsistent/misleading code indentation.
\item
  \textbf{cppcoreguidelines-avoid-goto}: Goto statements
\item
  \textbf{readability-magic-numbers}: Raw hex/decimal literals
  (e.g.~\texttt{0x401000})
\item
  \textbf{readability-redundant-casting}: Identifies unnecessary
  variable data/type casts (e.g.~\texttt{int\ x\ =\ (int)\ y;})
\item
  \textbf{cppcoreguidelines-pro-bounds-pointer-arithmetic}: Alerts on
  pointer arithmetic (e.g.~\texttt{ptr\ +\ 4} instead of
  \texttt{ptr{[}1{]}})
\end{enumerate}

In addition, we developed two custom \texttt{clang-tidy} checks to
identify:

\begin{enumerate}
\def\labelenumi{\arabic{enumi}.}
\tightlist
\item
  \textbf{qrs-function-pointer-cast}: Function pointer casts
  (e.g.~\texttt{void\ (*func)()\ =\ (void\ (*)())\ 0x401000;})
\item
  \textbf{qrs-hex-literal-cast}: Hex literal casts
  (e.g.~\texttt{int\ x\ =\ (int)\ 0x401000;})
\end{enumerate}

These patterns are common in decompiled code and reduce readability. The
more violations, the lower the idiomatic quality.

\subsection{Composite QRS Calculation and
Interpretation}\label{composite-qrs-calculation-and-interpretation}

\textbf{Full Formula}:

\begin{verbatim}
QRS = StructuralGate * CompositeScore

Where:
  StructuralGate =
    1.0 if BinarySimilarity >= 0.85
    else 0.1

  CompositeScore =
    0.34 * Normalize(LS) +
    0.33 * Normalize(SS) +
    0.33 * Normalize(IQ)

  Where Normalize(x) = clip(x, 0, 1)
\end{verbatim}

\textbf{QRS Thresholds}:

During our research and experiments, we found that the QRS threshold can
vary depending on the quality of the original source code and C
generated by the decompiler. However, we found that the following
thresholds provide some guidance:

\begin{itemize}
\tightlist
\item
  QRS \textgreater= 0.75: \textbf{Easy to understand and functionally
  equivalent}; suitable for security analysis or integration
\item
  QRS 0.55-0.75: \textbf{Readable but unlikely to be functionally
  equivalent}; requires improvements to both readability and functional
  equivalence
\item
  QRS 0.35-0.55: \textbf{Problematic}; significant readability issues
  and most likely contains compilation errors (resulting in a
  \texttt{0.1} functional equivalence score)
\item
  QRS \textless{} 0.35: \textbf{Unreadable}; the source code is unlikely
  to compile, will contain a large number of idiomatic violations,
  contain complex control flow structures, and use unfamiliar coding
  patterns often not found in human-written code.
\end{itemize}

Most code generated by a decompiler will start with a QRS score below
\texttt{0.20}. This is primarily because decompiler-generated C cannot
be re-compiled and because it usually contains many idiomatic
violations.

\textbf{Example QRS Reports}:

\textbf{Example 1: Function \texttt{compute\_hash}}

Metrics: Binary Similarity=0.94, LS=0.82, SS=0.68, IQ=0.71

\[
\begin{aligned}
\text{QRS} &= 1.0 \times (0.34 \times 0.82 + \\
  &\phantom{{}= 1.0 \times (}0.33 \times 0.68 + \\
  &\phantom{{}= 1.0 \times (}0.33 \times 0.71) \\
  &= 1.0 \times 0.74 = 0.74
\end{aligned}
\]

\textbf{Assessment}: Readable with cleanup

\textbf{Example 2: Function \texttt{buffer\_overflow\_safe\_copy}}

Metrics: Binary Similarity=0.92, LS=0.89, SS=0.91, IQ=0.88

\[
\begin{aligned}
\text{QRS} &= 1.0 \times (0.34 \times 0.89 + \\
  &\phantom{{}= 1.0 \times (}0.33 \times 0.91 + \\
  &\phantom{{}= 1.0 \times (}0.33 \times 0.88) \\
  &= 1.0 \times 0.89 = 0.89
\end{aligned}
\]

\textbf{Assessment}: Production-quality code

\section{Experimental Results and
Observations}\label{experimental-results-and-observations}

To evaluate the effectiveness of our QRS metric and iterative refinement
process we conducted two experiments on a synthetically generated
testing corpus of 210 simple C binaries (10,454 LoC across 12 unique
binaries) with varying levels of complexity, but always implementing
known software patterns (e.g.~reverse printing a string, bubble sort,
binary search, or linked lists). Both experiments leveraged a fully
automated workflow that ran refinement iterations per binary (max 16
parallel) until both the QRS and structural similarity thresholds were
met.

After completing a (simple, one-shot) Ghidra decompilation script to
generate the first version of the source code, each iteration would:

\begin{enumerate}
\def\labelenumi{\arabic{enumi}.}
\tightlist
\item
  Determine the structural similarity score using the `function-level
  similarity analysis using radare2's control-flow graph (CFG)
  comparison functionality' strategy.
\item
  Calculate each individual QRS sub-metric (Lexical Surprise, Structural
  Simplicity, and Idiomatic Quality) and the overall QRS score.
\item
  Generate improvement guidance based on the lowest scoring
  sub-metric(s).
\item
  Start a refinement iteration where the agent (ClaudeAgentSDK) is
  issued a prompt containing:

  \begin{itemize}
  \tightlist
  \item
    The QRS, structural similarity score, and the values for the QRS
    sub-metrics (Lexical Surprise, Structural Simplicity, and Idiomatic
    Quality)
  \item
    List of critical constraints (i.e.~``preserve functionality'',
    ``target weak metrics'', ``make incremental improvements'')
  \item
    The pre-compiled improvement guidance from step 3.
  \item
    Basic instructions on where to find and save source code.
  \end{itemize}
\end{enumerate}

The prompt did not contain detailed instructions as experimentation
showed that they are more likely to introduce small errors that are
amplified as iterations progressed. This is also the reason why the
dynamic prompt remained relatively small (\textasciitilde600 tokens) in
both experiments.

\subsection{Experiment 1: No command
execution}\label{experiment-1-no-command-execution}

In our first experiment we focussed on an LLM-only refinement process
(it did not grant the agent access to execute (Bash) commands) to
isolate our hypothesis and test the effectiveness of a metric and
iterative refinement strategy using an LLM.

\begin{figure}
\centering
\pandocbounded{\includegraphics[keepaspectratio,alt={Binary Refinement QRS Progress, each line represents 1 binary}]{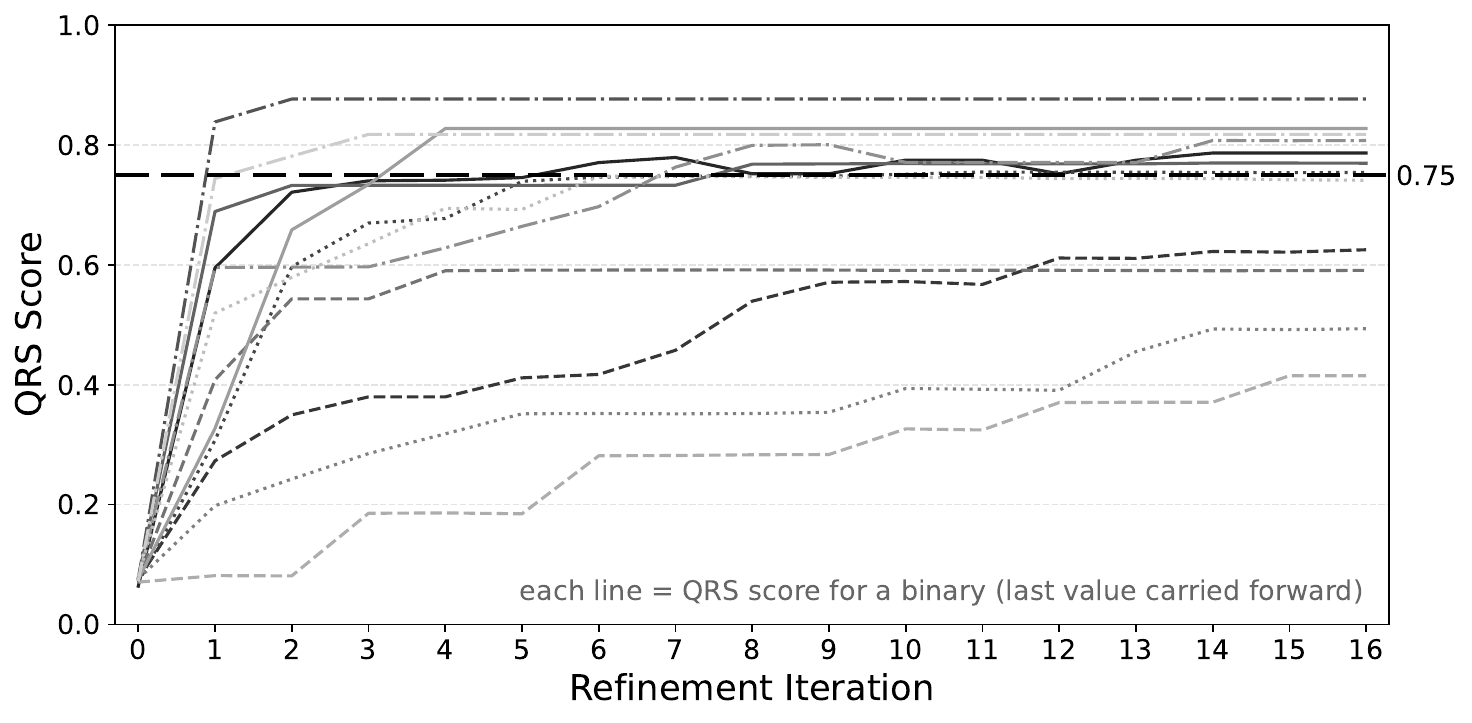}}
\caption{Binary Refinement QRS Progress, each line represents 1 binary}
\end{figure}

The experiment completed a total of 1453 iterations. 53 binaries
(25.24\%; 5.92 iterations per binary) failed to reach the QRS threshold
(0.75), and 21 binaries (23.33\%) failed to reach a structural
similarity score of 0.5. The average QRS improvement across all binaries
was +0.420. Structural similarity improved by +0.484. 0 iterations
resulted in a regression of QRS or structural similarity. This
demonstrates that the model correctly interpreted (and actioned) our
metrics, guidance, and instructions, confirming our hypothesis that the
model can effectively self-refine its outputs based on deterministic
feedback.

\begin{figure}
\centering
\pandocbounded{\includegraphics[keepaspectratio,alt={QRS and Sub-metric Distributions (no Bash)}]{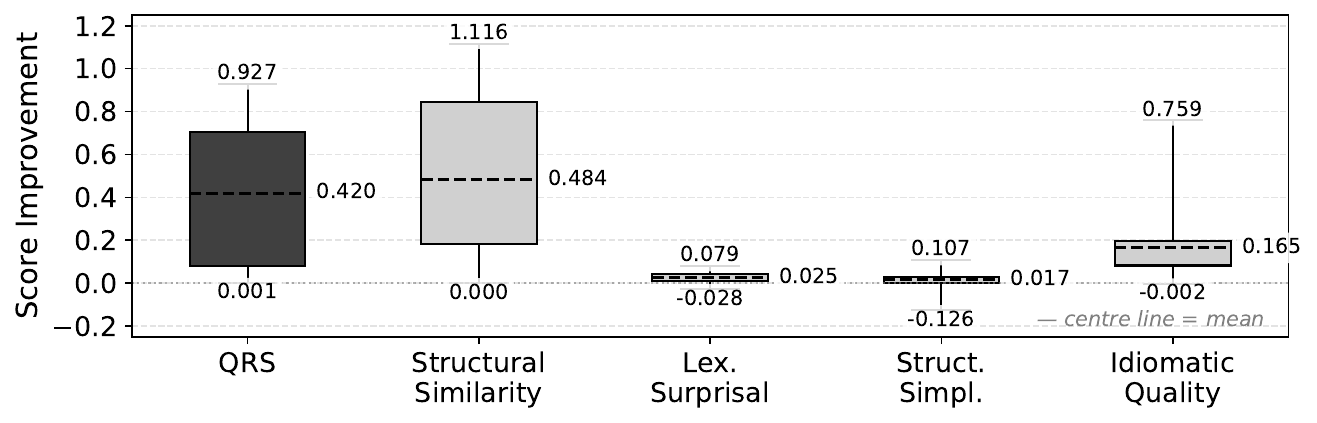}}
\caption{QRS and Sub-metric Distributions (no Bash)}
\end{figure}

Both `Lexical Surprisal' and `Structural Simplicity' experienced
isolated regressions. However, this remained contained to 1 or 2
iterations after which the metric course corrected back to an upward
trend.

\textbf{Observation}: \emph{We tested only accepting iterations that
resulted in a QRS increase by ignoring the output from a refinement
iteration if we observed a regression of the QRS metric. This approach
did not yield better results and appeared to trap the refinement process
in local maxima. Additionally, reviewing the source code produced by the
iteration indicated that QRS regression across one or two iterations can
yield a more substantial increase later.}

While the structural similarity score and the Idiomatic Quality
sub-metric were the primary contributors to the improvement of the QRS
metric it's important to note that all source code generated by the
decompiler produced source code that could not immediately be recompiled
without manual refinement to resolve the compiler errors. This meant
that the first iteration always started with a structural similarity
score of 0.1. Once the source code compiled, this score increased
significantly. Lexical Surprisal and Structural Simplicity saw much
smaller changes. This seems to indicate that the LLM correctly
interpreted the prompt and did not take any `moonshot' type approaches
during a refinement iteration. Had the model attempted to make large
sweeping changes to the codebase we would have expected to see large
regressions in these two metrics.

\textbf{Note:} \emph{`Idiomatic Quality' and `Structural Simplicity' are
expected to vary with real-world binaries. Because source code
`readability' is relative, and engineers are unlikely to produce source
code that produce idiomatic identical patterns with the same cyclomatic
complexity. So IQ and SS will likely need to be adjusted for different
binaries. Determining the optimal weighting for these sub-metrics could
be an area of future research.}

\subsection{Experiment 2: Command execution
enabled}\label{experiment-2-command-execution-enabled}

In our second experiment we provided the agent the ability to execute
commands using Bash, allowing it to run e.g.~\texttt{gcc} and review
compiler errors during a refinement iteration.

\begin{figure}
\centering
\pandocbounded{\includegraphics[keepaspectratio,alt={Binary Refinement QRS Progress}]{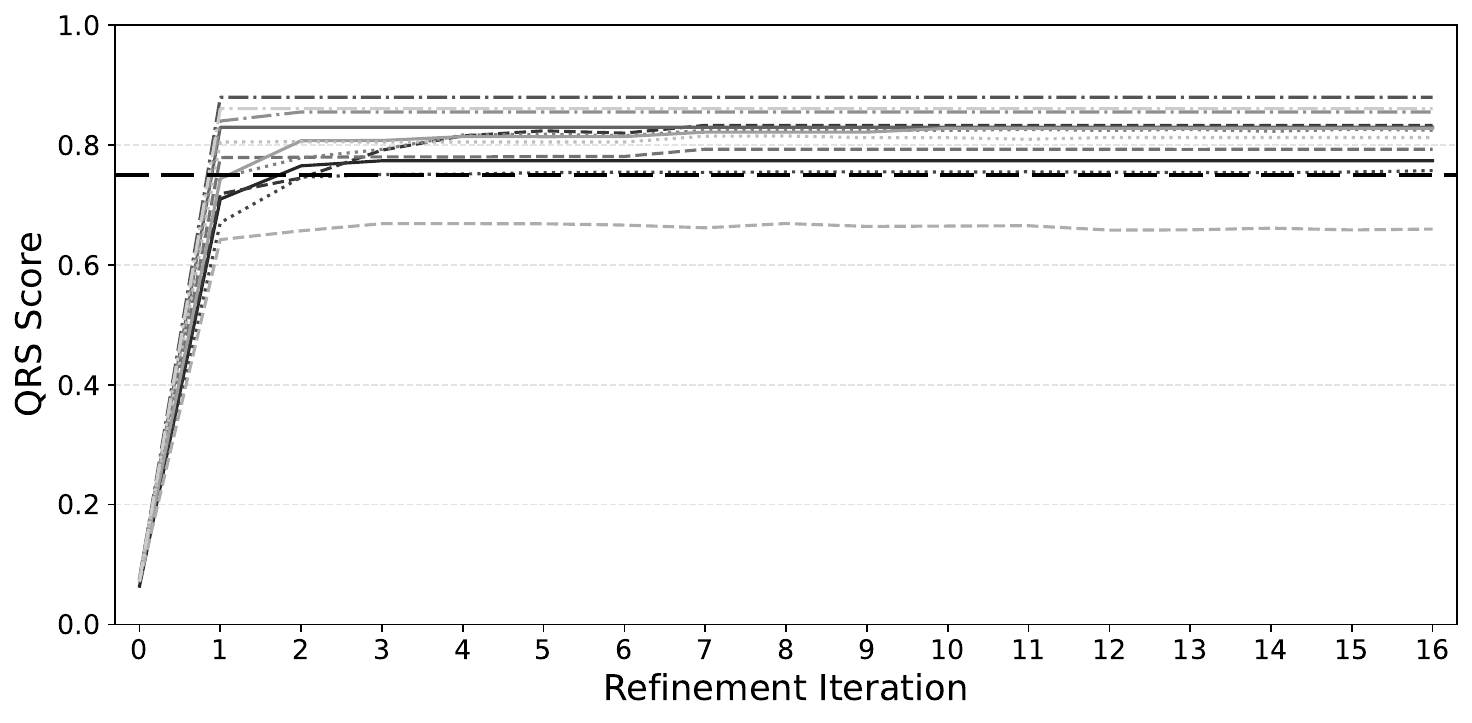}}
\caption{Binary Refinement QRS Progress}
\end{figure}

We observed a significant decrease in the number of iterations required
to reach the QRS and structural similarity thresholds. The experiment
saw a decrease in binaries failing (18) to reach the QRS and structural
similarity thresholds. This is an improvement of +34 binaries while also
needing fewer iterations (826; avg 2.933 iterations per binary). QRS
improved on average by +0.509 (+0.089) and structural similarity
improved by +0.582 (+0.098). The number of iterations that experienced a
regression in QRS or structural similarity remained at 0.

This data shows that providing the model with tools so it can validate
its own objectives (without providing it with the implementation details
of the QRS), the model was able to progress significantly faster.
However, the duration of each iteration saw a significant increase and
is unlikely to coincide with a reduced token consumption count.

\begin{figure}
\centering
\pandocbounded{\includegraphics[keepaspectratio,alt={QRS and Sub-metric Distributions (with Bash)}]{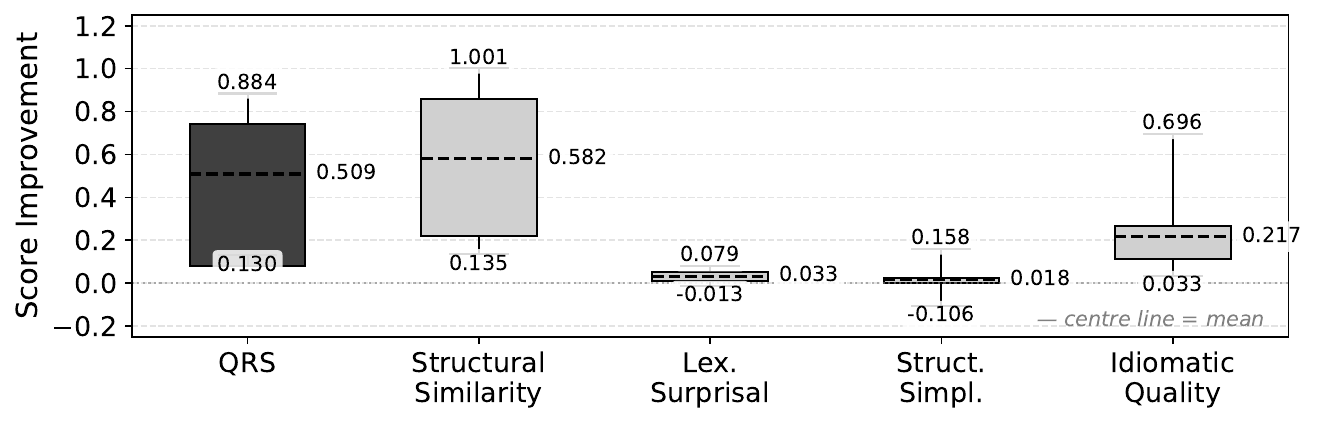}}
\caption{QRS and Sub-metric Distributions (with Bash)}
\end{figure}

\subsection{Results and Analysis}\label{results-and-analysis}

Both experiments demonstrate that:

\begin{itemize}
\tightlist
\item
  An LLM can convert decompiler-generated C into readable source code
  and resolve C compilation errors, without impacting the functional
  equivalence of the original binary.
\item
  Short iterations grounded in a deterministic objective-based metric
  enable LLMs to complete long-running and complex tasks without being
  sidetracked by context rot, hallucinations, premature completions, or
  instruction misprioritisation.
\item
  Providing an LLM with objective-aligned tools (e.g.~Bash) results in
  an average increase of 0.26 in QRS, and a decrease of 2.986 iterations
  to reach the QRS threshold.
\end{itemize}

\begin{figure}
\centering
\pandocbounded{\includegraphics[keepaspectratio,alt={Threshold and QRS Score Progress}]{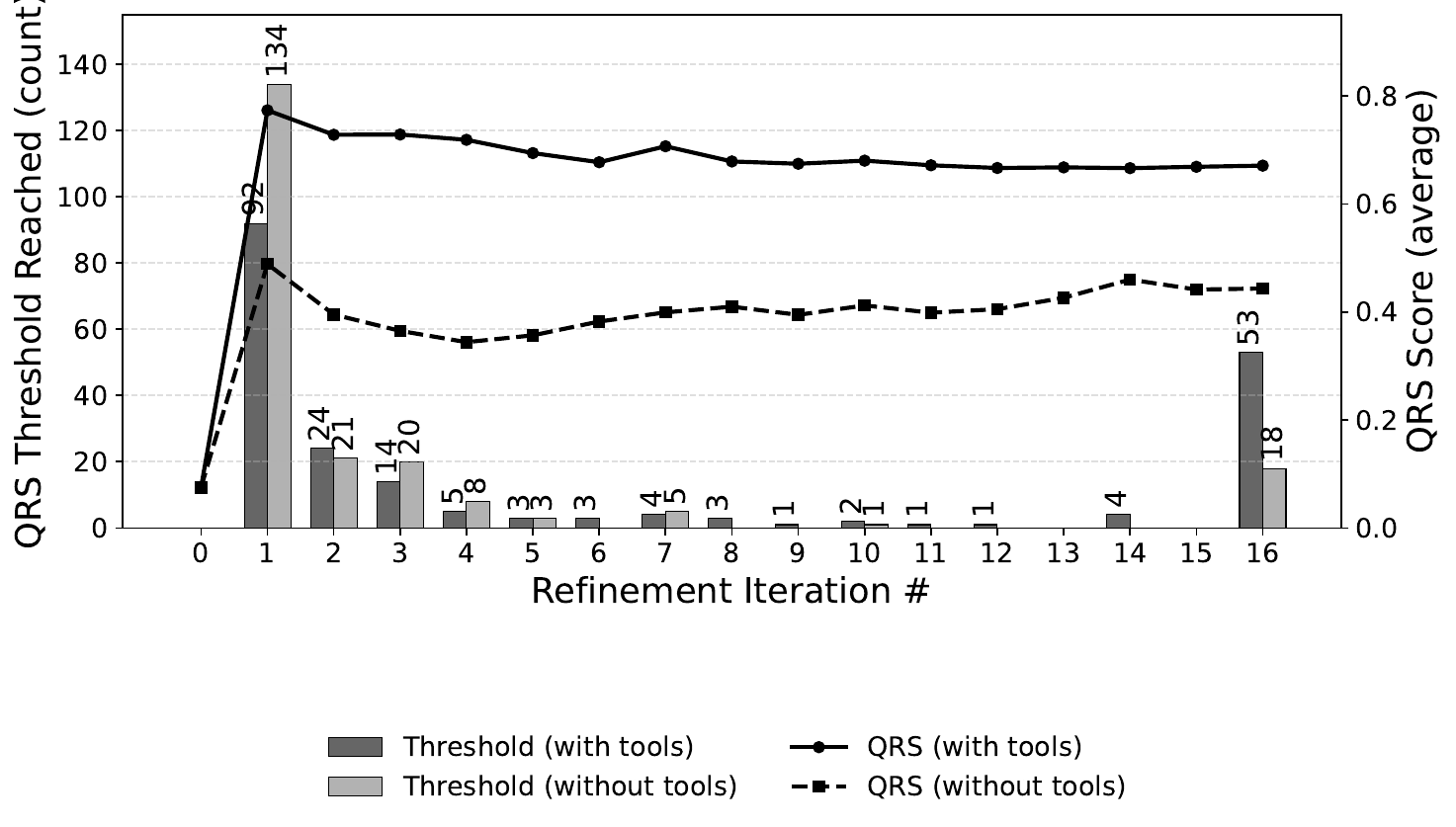}}
\caption{Threshold and QRS Score Progress}
\end{figure}

Our experiments also show that an LLM can more effectively improve
source code readability and structural similarity when it can validate
the prompt-based objectives using Bash. This is supported by the data
from our second experiment where we observed an average improvement of
+0.089 to the QRS, 0.098 to structural similarity, and 0.052 to
Idiomatic Quality.

\begin{figure}
\centering
\pandocbounded{\includegraphics[keepaspectratio,alt={Mean Improvement: No Bash vs With Bash}]{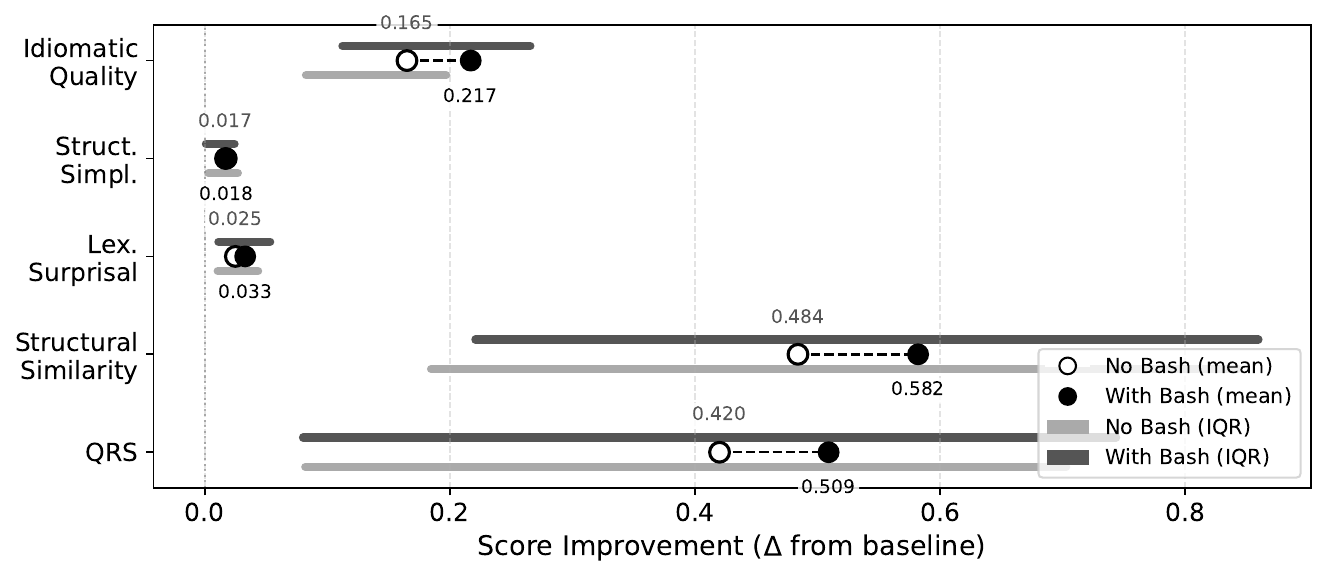}}
\caption{Mean Improvement: No Bash vs With Bash}
\end{figure}

Lexical Surprisal did not see a meaningful change. This is expected, as
it is a stabilising metric designed to detect hallucinations and
regressions in readability. Structural Simplicity also remained
relatively unchanged. This is expected to be the result of testing with
simple C binaries and is expected to be more volatile when applied to
more complex codebases with deeply nested control flow and data
structures.

\section{Future Work}\label{future-work}

This paper presents QRS-guided readability refinement, which operates on
decompiled code that has already achieved functional correctness
(\(\geq 0.85\) structural similarity). Our experimental evaluation
selected binaries where this prerequisite was achievable within
reasonable effort, allowing us to isolate and validate the QRS
framework's effectiveness for readability optimization. Below, we
outline both the out-of-scope prerequisite work and the research
directions that naturally extend from our findings.

\subsection{Decompilation Cleanup (Out of
Scope)}\label{decompilation-cleanup-out-of-scope}

QRS-guided refinement assumes it receives code that is already
functionally correct and compilable. Achieving this starting point, what
we term Phase 0, is itself a non-trivial problem that warrants dedicated
investigation.

\subsubsection{The Problem}\label{the-problem}

Binary decompilation is not deterministic. Automatic decompilers
frequently produce outputs with structural failures (broken jump tables,
incorrect control flow recovery, function boundary errors), type system
failures (pervasive void pointers, struct inference failures,
array/pointer confusion), and compiler or obfuscation artifacts. These
issues mean raw decompiler output often cannot compile, let alone match
the original binary's behaviour.

Phase 0 must resolve these fundamental quality issues before readability
refinement can begin. Without functional correctness, QRS scores become
meaningless: low scores reflect broken decompilation rather than
readability deficiencies, and agents waste iterations on cosmetic
changes that cannot address underlying structural problems.

\subsubsection{Why This Is Out of Scope}\label{why-this-is-out-of-scope}

The complexity of Phase 0 varies enormously based on binary
characteristics: stripped versus symboled binaries, obfuscated versus
straightforward code, and the quality of the decompiler itself. For
well-formed binaries, automated cleanup may take hours. For heavily
obfuscated or malformed binaries, the process can require weeks of
expert analysis. Cases involving virtualized malware, custom packers, or
deliberately corrupted headers could constitute independent research
contributions.

Our contribution lies in demonstrating that \emph{once functional
correctness is established}, multidimensional readability metrics can
guide systematic agent-driven improvement. Investigating the strategies,
tooling, and agent architectures needed to reliably achieve that
starting point is complementary but separate work.

\subsubsection{Research Directions for Phase
0}\label{research-directions-for-phase-0}

Future work on decompilation cleanup could explore:

\begin{itemize}
\tightlist
\item
  Agent-driven structural repair using MCP servers that expose
  decompiler APIs (Ghidra, IDA) for iterative CFG comparison, type
  propagation, and function signature recovery
\item
  Binary similarity as a cleanup signal, using radare2 CFG comparison to
  guide agents toward the \(\geq 0.85\) threshold through iterative
  structural fixes
\item
  Human-in-the-loop workflows for cases where automated approaches fail,
  particularly obfuscated binaries requiring domain expertise to correct
  control flow or annotate type information
\end{itemize}

\subsection{Extending the QRS
Framework}\label{extending-the-qrs-framework}

Several research directions follow naturally from the work presented in
this paper, with regard to the evolution of the propsed QRS Framework.

\subsubsection{Metric Refinement}\label{metric-refinement}

The current QRS weightings (LS: 0.34, SS: 0.33, IQ: 0.33) were
determined through empirical testing with equal-importance assumptions.
Future work could optimize these weights for specific domains.
Security-focused analysis may benefit from higher IQ weighting to
penalize obfuscation artifacts, while code comprehension tasks may
prioritize structural simplicity. Additionally, training domain-specific
language models for the Lexical Surprisal metric could improve
sensitivity to idioms particular to embedded systems, kernel code, or
protocol implementations.

\subsubsection{Broader Evaluation}\label{broader-evaluation}

Our evaluation corpus consists of 210 simple well-formed C binaries.
Evaluating QRS on additional languages (C++, Rust, Go), architectures
(ARM, MIPS), and binary complexity levels would strengthen confidence in
the framework's generalizability. Studying metric sensitivity and
robustness (how scores behave under adversarial conditions or at the
boundaries of decompiler capability) would help establish the
framework's limitations.

\subsubsection{Human Validation of QRS}\label{human-validation-of-qrs}

A limitation of this work is that QRS is used both to guide refinement
and to evaluate its effectiveness. This creates a circularity: an
improvement in QRS demonstrates that the agent optimized for QRS, but
does not independently confirm that the resulting code is more readable
to a human analyst. Code readability is inherently subjective and
influenced by analyst experience, domain familiarity, and task context.
QRS captures measurable proxies (familiar token patterns, low
complexity, absence of decompiler artifacts) but cannot fully encode the
subjective dimension of comprehension.

Validating that QRS improvements correlate with human readability
judgements is necessary to establish the framework's external validity.
This is out of scope for the present paper but represents a priority for
future work. Possible approaches include:

\begin{itemize}
\tightlist
\item
  Human evaluation studies: where participants rate before/after code
  samples and results are correlated with QRS deltas
\item
  Task-based benchmarks measuring whether higher QRS scores correspond
  to faster time-to-understanding or improved accuracy when analysts
  answer questions about function behaviour
\item
  Correlation with existing validated metrics such as CoReEval or other
  human-aligned readability assessments
\end{itemize}

Until such validation is performed, QRS should be understood as a
structured proxy for readability rather than a validated measure of
human comprehension.

\subsubsection{Integration with Reverse Engineering
Workflows}\label{integration-with-reverse-engineering-workflows}

The QRS framework currently operates as a standalone refinement loop.
Integrating it into larger reverse engineering workflows (combining
metric-guided cleanup with static analysis tools, symbolic execution,
and collaborative annotation systems) could produce end-to-end pipelines
that take a binary from raw decompilation through to analyst-ready
source code. This includes developing de-obfuscation pipelines that sit
adjacent to the readability refinement flow, handling the distinct but
complementary challenge of intentionally obscured code.

\section{Conclusion}\label{conclusion}

This paper documents a three-phase research journey in LLM
agent-assisted reverse engineering, culminating in a novel approach that
successfully produces readable, functionally correct decompiled code.

\subsection{Key Findings}\label{key-findings}

\begin{enumerate}
\def\labelenumi{\arabic{enumi}.}
\item
  \textbf{Un-guided Tool-Driven Steering Insufficient}: Providing agents
  with tools (Ghidra) without quantitative guidance leads to incomplete
  work and inconsistent quality. Agents need explicit, measurable
  objectives.
\item
  \textbf{Long-running agentic workflows with broad instructions are
  brittle}: Agents struggle to maintain focus and consistency over
  long-running workflows with high-level instructions. They require
  short feedback loops and clear, tasks that can be completed using an
  iterative process, and quantifiable goals to stay on task.
\item
  \textbf{Single-Metric Optimization is Vulnerable}: Using only
  structural similarity as a reward leads to metric gaming and
  unintended consequences. Agents rationally optimize around the metric,
  leading to high similarity but low readability.
\item
  \textbf{Multidimensional Metrics Enable Principled Refinement}:
  Combining three independent readability sub-metrics with a structural
  similarity gate creates a ``quality Pareto frontier'' that constrains
  agents from gaming. Each metric measures a distinct quality dimension.
\item
  \textbf{Dual Validation is Critical}: Combining QRS guidance with
  structural similarity validation ensures agents improve readability
  while maintaining correctness. The structural similarity gate makes
  equivalence non-negotiable.
\item
  \textbf{Decompilation Clean-up}: Decompilation to messy C, is more
  scalable than function-by-function tool-driven approaches. Agents then
  refine systematically.
\end{enumerate}

\subsection{Closing Remarks}\label{closing-remarks}

The journey from Phase 1 to Phase 3 reveals a fundamental principle: LLM
agents need rich, multidimensional guidance to solve complex tasks
without cutting corners. A single metric, even one measuring an
objectively important property, is insufficient. By measuring code
across multiple dimensions and combining metrics thoughtfully, we enable
agents to make principled refinements that improve readability while
maintaining correctness.

This work opens the door to larger questions about how to structure
rewards and metrics for agentic systems, particularly in domains where
correctness is non-negotiable but quality is subjective.\\
\strut \\
\strut \\
\strut \\
\strut \\
\strut \\
\strut \\
\strut \\
\strut \\
\strut \\
\strut \\
\strut \\
\strut \\
\strut \\
\strut \\
\strut \\
\strut \\
\strut \\
\strut \\
\strut \\
\strut \\
\strut \\
\strut \\
\strut \\
\strut \\
\strut \\
\strut \\
\strut \\
\strut \\
\strut \\
\strut \\
\strut \\
\strut \\

\end{multicols}

\section{Appendix A: References and Related
Work}\label{appendix-a-references-and-related-work}

\subsection{A.1 LLM-Assisted
Decompilation}\label{a.1-llm-assisted-decompilation}

{\def\LTcaptype{none} 
\begin{longtable}[]{@{}
  >{\raggedright\arraybackslash}p{(\linewidth - 2\tabcolsep) * \real{0.5071}}
  >{\raggedright\arraybackslash}p{(\linewidth - 2\tabcolsep) * \real{0.4929}}@{}}
\toprule\noalign{}
\begin{minipage}[b]{\linewidth}\raggedright
Reference
\end{minipage} & \begin{minipage}[b]{\linewidth}\raggedright
Description
\end{minipage} \\
\midrule\noalign{}
\endhead
\bottomrule\noalign{}
\endlastfoot
Tan et al.~(2024). ``LLM4Decompile: Decompiling Binary Code with Large
Language Models.'' EMNLP 2024. arXiv:2403.05286 & First open-source LLM
series (1.3B--33B) trained for decompilation; introduces end-to-end and
Ghidra-refinement approaches \\
Zou et al.~(2025). ``D-LiFT: Improving LLM-based Decompiler Backend via
Code Quality-driven Fine-tuning.'' arXiv:2506.10125 & Multi-dimensional
``D-Score'' for decompiled code quality guiding RL fine-tuning;
preserves accuracy while improving readability \\
Wong et al.~(2023). ``Refining Decompiled C Code with Large Language
Models.'' arXiv:2310.06530 & LLM augmentation of IDA-Pro output
achieving 75\% recompilation success; precursor to readability-focused
refinement \\
Zhang et al.~(2026). ``CoDe-R: Refining Decompiler Output with LLMs via
Rationale Guidance and Adaptive Inference.'' IJCNN 2026.
arXiv:2604.12913 & Two-stage refinement addressing logical
hallucinations via semantic cognitive enhancement and hybrid
verification \\
Zhang et al.~(2026). ``Constraint-Guided Multi-Agent Decompilation for
Executable Binary Recovery.'' arXiv:2604.23940 & Multi-agent LLM
approach with hierarchical validation (syntactic, compilation,
behavioral equivalence); achieves 84--97\% re-executability \\
Tan et al.~(2025). ``SK2Decompile: LLM-based Two-Phase Binary
Decompilation from Skeleton to Skin.'' arXiv:2509.22114 & Decomposes
decompilation into structure recovery and identifier naming as
independent phases \\
Wang et al.~(2025). ``Context-Guided Decompilation: A Step Towards
Re-executability (ICL4Decomp).'' arXiv:2511.01763 & In-context learning
guiding LLMs toward re-executable decompilation; \textasciitilde40\%
improvement over prior state-of-the-art \\
Shypula et al.~(2026). ``Decaf: Improving Neural Decompilation with
Automatic Feedback and Search.'' arXiv:2605.11501 & Compiler
feedback-driven iterative improvement (26\% to 83.9\% success);
demonstrates automatic feedback outperforms additional training data \\
Dramko et al.~(2025). ``Idioms: Neural Decompilation With Joint Code and
Type Definition Prediction.'' arXiv:2502.04536 & Recovers idiomatic code
constructs during decompilation through joint prediction of code and
type definitions \\
Enders et al.~(2023). ``dewolf: Improving Decompilation by Leveraging
User Surveys.'' BAR Workshop 2023. arXiv:2205.06719 & Human-validated
decompilation quality improvements; motivates automated readability
metrics \\
\end{longtable}
}

\subsection{A.2 Code Readability Metrics and Automated
Assessment}\label{a.2-code-readability-metrics-and-automated-assessment}

{\def\LTcaptype{none} 
\begin{longtable}[]{@{}
  >{\raggedright\arraybackslash}p{(\linewidth - 2\tabcolsep) * \real{0.4716}}
  >{\raggedright\arraybackslash}p{(\linewidth - 2\tabcolsep) * \real{0.5284}}@{}}
\toprule\noalign{}
\begin{minipage}[b]{\linewidth}\raggedright
Reference
\end{minipage} & \begin{minipage}[b]{\linewidth}\raggedright
Description
\end{minipage} \\
\midrule\noalign{}
\endhead
\bottomrule\noalign{}
\endlastfoot
Halstead, M. H. (1977). ``Elements of Software Science'' & Foundational
work on quantitative code metrics \\
McCabe, T. J. (1976). ``A Complexity Measure.'' IEEE TSE 2(4) &
Cyclomatic complexity measurement via control flow graph analysis \\
Ray et al.~(2015). ``On the Naturalness of Buggy Code.''
arXiv:1506.01159 & Establishes that buggy/unnatural code has higher
entropy; foundational motivation for lexical surprisal as a readability
signal \\
Ouedraogo et al.~(2025). ``Human-Aligned Code Readability Assessment
with Large Language Models (CoReEval).'' arXiv:2510.16579 & First
large-scale benchmark for LLM-based readability assessment (1.4M
evaluations, 10 LLMs); validates LLMs as readability judges \\
Horikawa et al.~(2026). ``Do AI Agents Really Improve Code
Readability?'' arXiv:2603.13723 & Finds readability-focused agent
commits often degrade traditional quality metrics; motivates
multi-metric approaches like QRS \\
Liu et al.~(2025). ``On Iterative Evaluation and Enhancement of Code
Quality Using GPT-4o (CodeQUEST).'' arXiv:2502.07399 & LLM-based
iterative multi-dimensional code quality optimization achieving 52.6\%
mean improvement across readability, maintainability, efficiency, and
security \\
Zheng et al.~(2024). ``Beyond Correctness: Benchmarking
Multi-dimensional Code Generation (RACE).'' arXiv:2407.11470 &
Multi-dimensional evaluation framework (readability, maintainability,
correctness, efficiency); demonstrates correctness-only metrics are
insufficient \\
Midolo et al.~(2026). ``From Human to Machine Refactoring: Assessing
GPT-4's Impact on Python Class Quality and Readability.''
arXiv:2601.13139 & Finds GPT-4o reduces code smells at the cost of
decreased readability; validates the need for joint optimization across
quality dimensions \\
\end{longtable}
}

\subsection{A.3 Binary Similarity and Semantic
Equivalence}\label{a.3-binary-similarity-and-semantic-equivalence}

{\def\LTcaptype{none} 
\begin{longtable}[]{@{}
  >{\raggedright\arraybackslash}p{(\linewidth - 2\tabcolsep) * \real{0.6038}}
  >{\raggedright\arraybackslash}p{(\linewidth - 2\tabcolsep) * \real{0.3962}}@{}}
\toprule\noalign{}
\begin{minipage}[b]{\linewidth}\raggedright
Reference
\end{minipage} & \begin{minipage}[b]{\linewidth}\raggedright
Description
\end{minipage} \\
\midrule\noalign{}
\endhead
\bottomrule\noalign{}
\endlastfoot
Radare2 project (radare.org) & Native CFG comparison tool for binary
diffing used in our structural similarity gate \\
Saul et al.~(2024). ``Is Function Similarity Over-Engineered? Building a
Benchmark.'' arXiv:2410.22677 & Evaluates binary function similarity
approaches; contextualizes CFG-based validation \\
Zuo et al.~(2024). ``BinSimDB: Benchmark Dataset Construction for
Fine-Grained Binary Code Similarity Analysis.'' arXiv:2410.10163 &
Benchmark infrastructure for binary similarity tool evaluation \\
\end{longtable}
}

\subsection{A.4 Agent-Based Code
Transformation}\label{a.4-agent-based-code-transformation}

{\def\LTcaptype{none} 
\begin{longtable}[]{@{}
  >{\raggedright\arraybackslash}p{(\linewidth - 2\tabcolsep) * \real{0.5399}}
  >{\raggedright\arraybackslash}p{(\linewidth - 2\tabcolsep) * \real{0.4601}}@{}}
\toprule\noalign{}
\begin{minipage}[b]{\linewidth}\raggedright
Reference
\end{minipage} & \begin{minipage}[b]{\linewidth}\raggedright
Description
\end{minipage} \\
\midrule\noalign{}
\endhead
\bottomrule\noalign{}
\endlastfoot
Yang et al.~(2024). ``SWE-agent: Agent-Computer Interfaces Enable
Automated Software Engineering.'' arXiv:2405.15793 & Foundational work
on LLM agents for software engineering; demonstrates interface design
significantly impacts agent performance \\
Zhang et al.~(2024). ``AutoCodeRover: Autonomous Program Improvement.''
arXiv:2404.05427 & Autonomous agent combining code search with iterative
patching for program improvement \\
Yuksel \& Sawaf (2024). ``A Multi-AI Agent System for Autonomous
Optimization via Iterative Refinement and LLM-Driven Feedback Loops.''
arXiv:2412.17149 & Multi-agent autonomous optimization through iterative
feedback; parallels metric-driven refinement workflows \\
Hou \& Yang (2026). ``AI Agent for Reverse-Engineering Legacy
Finite-Difference Code and Translating to Devito.'' arXiv:2601.18381 &
LangGraph-based agent with multi-stage iterative workflows for code
transformation with quality-based refinement \\
\end{longtable}
}

\subsection{A.5 Reward Design and Multi-Objective
Optimization}\label{a.5-reward-design-and-multi-objective-optimization}

{\def\LTcaptype{none} 
\begin{longtable}[]{@{}
  >{\raggedright\arraybackslash}p{(\linewidth - 2\tabcolsep) * \real{0.4708}}
  >{\raggedright\arraybackslash}p{(\linewidth - 2\tabcolsep) * \real{0.5292}}@{}}
\toprule\noalign{}
\begin{minipage}[b]{\linewidth}\raggedright
Reference
\end{minipage} & \begin{minipage}[b]{\linewidth}\raggedright
Description
\end{minipage} \\
\midrule\noalign{}
\endhead
\bottomrule\noalign{}
\endlastfoot
Goodhart, C. A. (1975). ``Problems of Monetary Management: The U.K.
Experience'' & Goodhart's Law: when a measure becomes a target, it
ceases to be a good measure \\
Hadfield-Menell et al.~(2016). ``Cooperative Inverse Reinforcement
Learning.'' arXiv:1606.03137 & Multi-objective reward specification for
cooperative AI systems \\
Zhou et al.~(2025). ``From Obfuscated to Obvious: A Comprehensive
JavaScript Deobfuscation Tool.'' arXiv:2512.14070 & Multi-dimensional
evaluation metrics including LLM-based readability for deobfuscated
code; achieves 4x readability improvement \\
\end{longtable}
}

\section{Appendix B: Glossary}\label{appendix-b-glossary}

{\def\LTcaptype{none} 
\begin{longtable}[]{@{}
  >{\raggedright\arraybackslash}p{(\linewidth - 2\tabcolsep) * \real{0.2778}}
  >{\raggedright\arraybackslash}p{(\linewidth - 2\tabcolsep) * \real{0.7222}}@{}}
\toprule\noalign{}
\begin{minipage}[b]{\linewidth}\raggedright
Term
\end{minipage} & \begin{minipage}[b]{\linewidth}\raggedright
Definition
\end{minipage} \\
\midrule\noalign{}
\endhead
\bottomrule\noalign{}
\endlastfoot
\textbf{QRS} & Quantitative Readability Score; composite metric
measuring code readability \\
\textbf{LS} & Lexical Surprisal; code familiarity via language model
perplexity \\
\textbf{SS} & Structural Simplicity; control flow and function
complexity \\
\textbf{IQ} & Idiomatic Quality; adherence to common coding patterns and
idioms \\
\textbf{Structural Similarity Gate} & Quality gate ensuring structural
similarity (\(\geq 0.85\)) before computing QRS \\
\textbf{radare2 CFG comparison} & Binary similarity validation using
radare2's control flow graph comparison \\
\textbf{MCP} & Model Context Protocol; tool interface for LLMs \\
\end{longtable}
}

\begin{multicols}{2}

\end{multicols}

\end{document}